\documentclass[american,aps,prl,reprint,showpacs,superscriptaddress,altaffilletter]{revtex4-2}
\usepackage{lmodern}
\usepackage{lmodern}
\usepackage[T1]{fontenc}
\usepackage[utf8]{inputenc}
\usepackage{babel}
\usepackage{amsmath}
\usepackage{amssymb}
\usepackage{graphicx}
\usepackage{wasysym}
\usepackage{ulem}
\usepackage{siunitx}
\usepackage[unicode=true,
 bookmarks=true,bookmarksnumbered=false,bookmarksopen=false,
 breaklinks=false,pdfborder={0 0 0},pdfborderstyle={},backref=false,colorlinks=true]
 {hyperref}
\hypersetup{
 pdfborderstyle={},pdfborderstyle={},pdfborderstyle={},pdfborderstyle={},pdfborderstyle={},pdfborderstyle={},linkcolor=blue,citecolor=blue}

\makeatletter
\usepackage{comment}

\makeatother

\begin{document}
%\title{Tuning the restitution coefficient  of closed containers partially filled with water}
%\title{Fluid motion for reducing the bounce of partially filled containers}

%\title{Water Redistribution, Jet Sloshing, and Landing Angle Rule the Bottle Flip Challenge}

%\title{Water Motion and the Playfulness of the Bottle Flip Challenge}
\title{Fluid Motion Makes the Bottle-Flip Challenge Mechanically Unintuitive But Viable}

\author{Patricio Morales}
\affiliation{Instituto de Ciencias de la Ingeniería, Universidad de O'Higgins,
Av. Libertador Bernardo O'Higgins 611, Rancagua, Chile}

\author{Victor Ahumada}
\affiliation{Instituto de Ciencias de la Ingeniería, Universidad de O'Higgins,
Av. Libertador Bernardo O'Higgins 611, Rancagua, Chile}

\author{Leonardo Gordillo}
\email{leonardo.gordillo@usach.cl}
\affiliation{Departamento de Física, Facultad de Ciencia, Universidad de Santiago de Chile, ~\\
 Av. Víctor Jara 3493, Estación Central, Santiago, Chile}
 
 \author{Pablo Guti\'errez}
\email{pablo.gutierrez@uoh.cl}
\affiliation{Instituto de Ciencias de la Ingeniería, Universidad de O'Higgins,
Av. Libertador Bernardo O'Higgins 611, Rancagua, Chile}

\begin{abstract} %OJO que ahora PRL pide abstract < 600 CARACTERES

The bottle-flip challenge \textemdash the upright landing of a partially filled bottle after tossing and flipping it in the air\textemdash  unexpectedly became a viral mechanics exercise. Through high-speed visualization, we evidence that fluid content strongly affects the challenge mechanics at every stage. First, fluid motion hinders bottle rotation through water redistribution after release and, later, through jet impact during free flight. Water also attenuates the bounce at touchdown, which is crucial for challenge success. The fluid makes the mechanics unintuitive and yet builds up the challenge feasibility.
\end{abstract}

\maketitle

The early stages of human life are an endless exploration of the mechanics of the surroundings. Toddlers are known for their eagerness to learn mechanics by tossing, pouring, bouncing, or splashing quotidian objects \cite{PsicologiaPiaget_1966}. Powered by hand-in-hand integration of sensory stimuli with coarse and fine motor skills, child mechanical exploration runs in parallel with the refinement of cognitive experimental skills \cite{KhapardeShaker_2020,HumanMotorDevelopment}. Concepts as repeatable, controllable, predictable, or unexpected become part of the interpretation and classification of physical-world outcomes. Mechanical exploration, however, plunges dramatically as we age \cite{Branta_etAl_1984}. After all that childhood experimentation, discovering new “moves” becomes unlikable as we grow, especially when the moves involve simple everyday objects.

The bottle-flip challenge video, which went viral in 2016, accidentally threw millions of people worldwide into tossing, flipping, and landing upright a bottle partially filled with water \cite{News1_Link,News2_Link}. What made this mechanical trick, among many others available, so appealing to the general audience? Consider first that dropping an empty bottle, even straight from rest,  leads to erratic bounces back and rarely to straight-down landing  \cite{Andrade_2023}. So, flipping an empty bottle to make it land upright after completing a single turn on air would require robotic precision. Filling the bottle with some water may look like raising the bar in mechanical terms, as the motion of partially filled objects is even harder to anticipate (e.g., readers could think of the forces they experience when carrying filled buckets).
In a nonsensical turn of events, a person with average mechanical skills and after some practice can master the technique of flipping and releasing a partially filled bottle so it could end up at an abrupt and stable stop in its upright position. 

Aiming at understanding the popularity of the bottle-flip challenge, this article scrutinizes a three-fold physical process at its root that ultimately determines the outcome: (1) an initial redistribution of water after release, with implications in angular momentum conservation \cite{DekkerEtAl_2018}; (2) a sloshing during the free flight of the bottle, and (3) the stomping of the bottle at landing, due to the abrupt descend of water \cite{Andrade_2023}. We explain how the counterintuitive role of hydrodynamics is what makes the challenge not only enticing but also viable.

\begin{figure}[b]%[htbp]
\begin{centering}
\includegraphics[width=1\columnwidth]{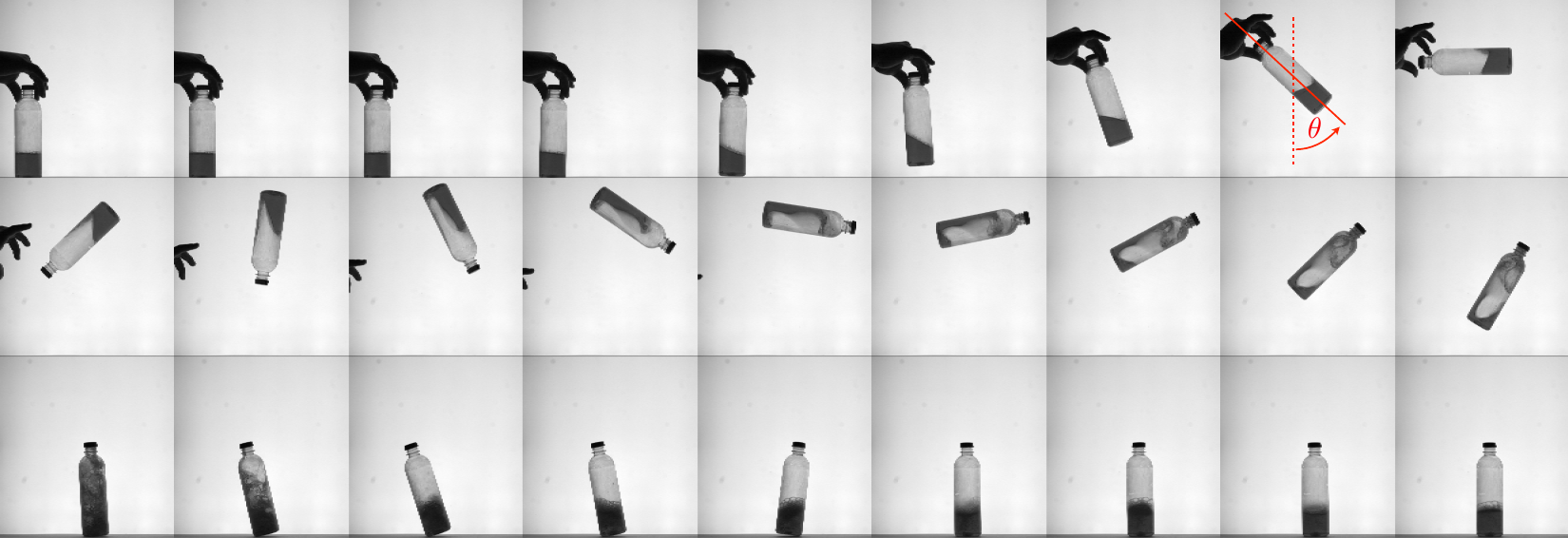}
\par\end{centering}
\caption{Successful execution of the bottle flip challenge. The top row shows the performer's action to set the motion, the middle row shows the bottle's free flight, and the bottom row shows the landing. The top and middle rows have time steps of 37.5~ms, while the bottom row uses time steps of 112.5~ms. The bottle filling volume fraction is $\phi = 0.33$. \label{Fig:FlipSequence}}
\end{figure}

To start our analysis, we focus on a successful completion of the bottle-flip challenge. The time-lapse of Fig.~\ref{Fig:FlipSequence} shows the execution stages in detail.  The upper row shows two necessary gestures to spin the bottle: turning from rest to releasing. During its flight, the bottle moves freely (middle row), and so does the water inside. Indeed, one can notice the spreading of water along the bottle and the sloshing of a jet during this freely moving stage. Finally, the bottle hits the ground and lands (bottom row). The touchdown reveals
 the challenge outcome. Successful executions of the challenge end up with the bottle slightly tumbling before ending upright. Unsuccessful ones end with an improper bounce or a sudden tip-over.

\begin{figure}[t]
\begin{centering}
\includegraphics[width=1\columnwidth]{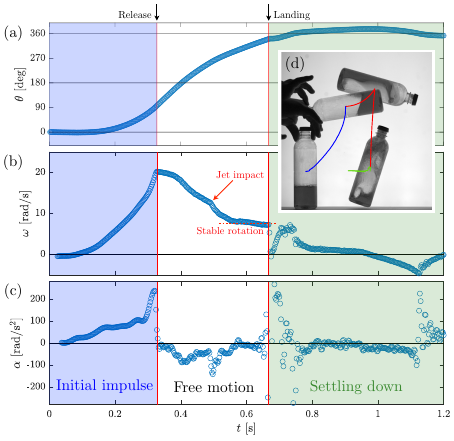}
\par\end{centering}
\caption{Bottle kinematics for the challenge execution in Fig.~\ref{Fig:FlipSequence}. (a) We measured the bottle's orientation $\theta$ regardless of the water inside. $\theta$ is defined in Fig.~\ref{Fig:FlipSequence}. (b) Angular velocity of the bottle $\omega$, and (c) angular acceleration $\alpha$, calculated as the time derivatives of $\theta$ and $\omega$, respectively. Three motion stages are emphasized with background colors: blue for the initial impulse and green for the stage of settling down. In free flight, we emphasize the jet sloshing and a stable, non-zero value for $\omega$. The inset (d) presents the bottle's trajectory together with four frames: an initial one at $t = 0$~ms; another at release ($t_r = 330$~ms); one at jet sloshing ($t_j = 490$~ms); and the final one, just after landing ($t_l =675$~ms).\label{Fig:ThetaOmega}}
\end{figure}

To measure the kinematics of the bottle during the challenge, we performed experiments and video-sequenced them for thorough analysis. We used cylindrical PET bottles ($D$ = 6 cm \diameter~ and $h=$ 18 cm height, as in \cite{Andrade_2023}). We record videos using a Phantom camera (VEO 410S) at 400~Hz and analyze data with Matlab\textregistered. We lightly sanded the bottle surface to make it opaque and ease the detection of the water-content motion.
%handle each frame depending on whether the bottle or the water content motion is in our target. 
We use high-threshold binarized images to track only the bottle's dynamics and grayscale raw images to examine the fluid motion inside. This results in two different and complementary analyses of the same videos. For bottle-targeted analysis, we extract in each frame the bottle's rigid-part centroid ($x_b$,$y_b$), which is independent of the water distribution; together with its orientation $\theta$ as defined in Fig.~\ref{Fig:FlipSequence}. The resultant signal $\theta (t)$  is low-pass filtered at 70~Hz, while their derivatives $\omega (t)$ and $\alpha (t)$ are obtained digitally as shown in Fig.~\ref{Fig:ThetaOmega}. 

%Using a procedure adapted from (refs), we estimated the position of the center of mass (see Supplemental Material for details). In Fig.~2, we included an example, together with the evolution of the centroid's trajectory. 

\begin{figure}[t!]%[htbp]
\begin{centering}
\includegraphics[width=1\columnwidth]{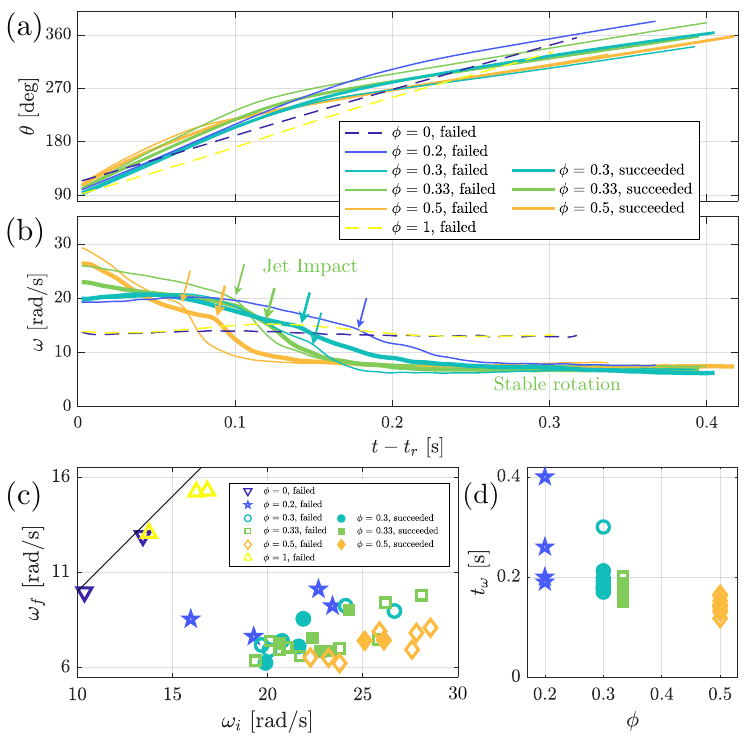}
\par\end{centering}
\caption{Bottle free-flight motion. As in Fig.~\ref{Fig:ThetaOmega}, we present (a) orientation $\theta$ and (b) angular velocity $\omega$, regardless of the water inside. The time is measured from the release $t_r$ to the landing. Color represents the filling volume fraction $\phi$, going from dark to light colors when the water content increases. The control cases are the full bottle ($\phi = 1$, yellow dashed lines) and the empty bottle ($\phi = 0$, dark blue dashed lines). Arrows indicate instants at which jet sloshing occurs. For (a) and (b), thin lines represent failed attempts, and thicker lines, successful ones. (c) Final angular velocity $\omega_f$ versus initial one $\omega_i$. The black line indicates constant $\omega$. (d) Time at which $\omega$ reaches a stable value versus filling volume fraction $\phi$. For (c) and (d), failed throwings are drawn with open symbols and successful ones with filled symbols. \label{Fig:FreeMotion}}
\end{figure}

Figure~\ref{Fig:ThetaOmega} shows a kinematic angular characterization of the bottle in a successful execution of the challenge. 
Although the orientation angle $\theta$ gently evolves to complete a turn [see Fig.~\ref{Fig:ThetaOmega}(a)], both angular velocity $\omega$ [Fig.~\ref{Fig:ThetaOmega}(b)] and angular acceleration $\alpha$ [Fig.~\ref{Fig:ThetaOmega}(c)] reveal abrupt changes between the three main stages of motion. During the initial impulse, the signal of $\alpha$ demonstrates that the thrower accelerates the bottle until releasing it with a final sharp impulse. Then, during free flight, $\omega$ shows that the bottle's rotation speeds down until it reaches a nearly stable non-zero rotational velocity. Below, we shall discuss how water motion is responsible for bottle rotation deceleration. Finally, after the bottle lands, it still takes some time to settle in its upright final state. 
%Although the orientation angle $\theta$ gently evolves to complete a turn [see Fig.~\ref{Fig:ThetaOmega}(a)], the angular velocity $\omega$ [Fig.~\ref{Fig:ThetaOmega}(b)] reveals abrupt changes between the three main stages of motion. During the initial impulse, the rate-of-change $\omega$ demonstrates that the thrower accelerates the bottle until releasing it with a final sharp impulse. Then, during free flight, $\omega$ shows that the bottle's rotation speeds down until it reaches a nearly stable non-zero rotational velocity. Below, we shall discuss how water motion is responsible for bottle rotation deceleration. Finally, after the bottle lands, it still takes some time to settle in its upright final state. 
We present the trajectory of the bottle's rigid-part centroid [$x_b(t)$,$y_b(t)$] in Fig.~\ref{Fig:ThetaOmega}(d), superposed on a set of other four frames (see caption). While we observed the expected abrupt changes in the trajectory during release and landing (highlighted with a different color), the trajectory exhibits another, less expected, sharp change during its free flight. 

\begin{figure*}[tb]
\begin{centering}
\includegraphics[width=2\columnwidth]{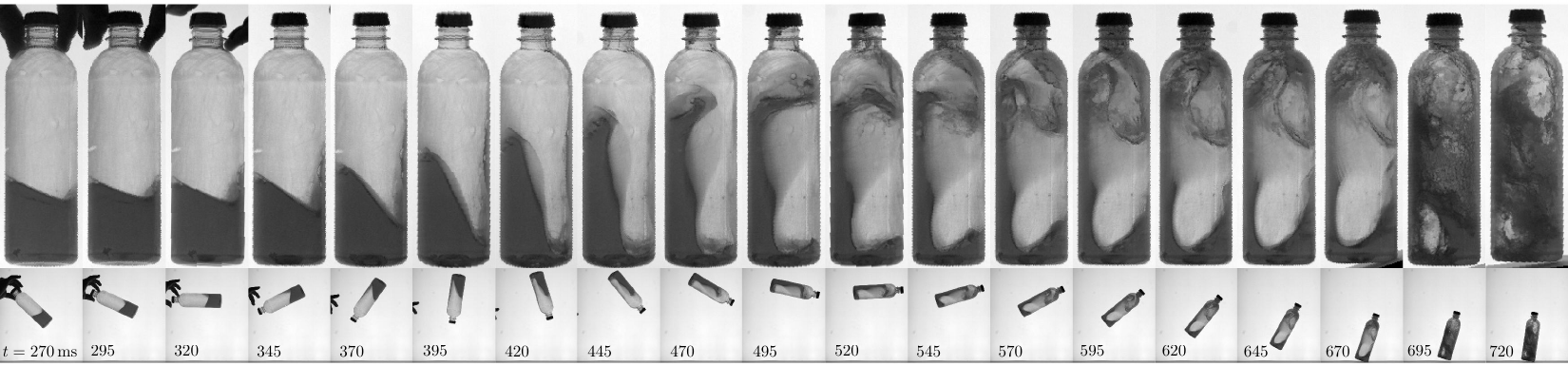}
\par\end{centering}
\caption{Water motion inside the bottle. The top panels present the same sequence as in Fig.~\ref{Fig:FlipSequence} but in the bottle's comoving and corotation frame of reference, as provided by $x_b(t)$, $y_b(t)$ and $\theta (t)$. The bottom panels are the corresponding frames in the laboratory frame of reference. Time (in milliseconds) is indicated in the bottom panels. This representation sets apart the water dynamics inside the bottle during the execution of the challenge.}\label{Fig:RotReferential}
\end{figure*}

%\paragraph{Free-flight motion}

%As every training process involves learning from mistakes, unsuccessful bottle-flip challenge executions may still disclose relevant motion features.  We performed thousands of experiments and video-sequenced them regardless of their outcome. The large sample allows us to sweep control parameters on the range of an average-person skill and simultaneously collect statistics. Among the countless experiments we performed and video-sequenced, we did not discard our failed executions, providing further statistics. Hence, the data shown below includes challenge executions regardless of their outcome.   
Inspired by the fact that training benefits from repeating and learning from mistakes, we significantly extended the sample size of our experimental runs over two hundred, video-sequencing and processing each realization regardless of its outcome, i.e., failed executions were not discarded but tagged.  Our sample size is large enough to collect statistics on the span of control parameters within the range of an average-person skill. We also included experiments with several filling volume fractions $\phi$ ranging from a brimful bottle, $\phi = 1$, to an empty one, $\phi = 0$. Both $\phi=1$ and $\phi=0$ serve as control cases for comparison as they preclude motion transfer to the fluid content.  
Figure~\ref{Fig:FreeMotion} depicts the kinematics of several representative bottle-flip challenge trials during their free flight, adequately tagged according to their outcome.
The evolution curves start from the release time, $t_r$, and evolve until landing, which happens at differing times. The first piece of information from Fig.~\ref{Fig:FreeMotion}(a)-(b) is that the kinematics between succeeded and failed launches hardly differ. The only exceptions are the two control cases, brimful and empty bottle, whose dynamics are different even qualitatively, and for which we never managed to accomplish a straight landing despite our many attempts. 

Setting aside control cases, we generally observe that as the bottle rotates, Fig.~\ref{Fig:FreeMotion}(a) shows a monotonous increase in $\theta$ until the touchdown. The orientation ends widely, from 3/4~turn ($\ang{270}$) to 5/4~turn ($\ang{450}$). The angular velocity $\omega$ is also rich in information. Indeed, curves $\omega$ vs $t$ in Fig.~\ref{Fig:FreeMotion}(b) show an initial value $\omega_i$ that decays until reaching a stable constant value $\omega_f$, as previously reported in \cite{DekkerEtAl_2018}. Figure~\ref{Fig:FreeMotion}(c) emphasizes that $\omega_i$ span is much larger in magnitude and dispersion than $\omega_f$'s, while the control cases ($\phi$ = 0 and $\phi$ = 1) clearly outlie from the trend as $\omega$ remains almost constant during flight. 
The bottle reaches its stable rotation $\omega_f$ after a time $t_{\omega}$, shown in Fig.~\ref{Fig:FreeMotion}(d). Also, $t_{\omega}$ shortens with increasing volume of water, as it has less empty space to move along the bottle's inside.
%While noisier, angular acceleration reveals extra features [see Fig.~\ref{Fig:FreeMotion}(c)]. 
Many curves in Fig.~\ref{Fig:FreeMotion}(b) show the same two acceleration events highlighted in Fig.~\ref{Fig:ThetaOmega}(c): the first acceleration event is the simple decrease in $\omega$ after hand release, while the second event correlates with jet sloshing. 
%Otherwise, more substantial accelerations are produced with larger volumes of water, as quantified by $\phi$ [see Fig.~\ref{Fig:FreeMotion}(g)]. 
After $t_{\omega}$, $\omega$ remains close to a constant value (stable rotation) until the landing.
%\textcolor{blue}{
In summary, the evolution of the orientation $\theta$ and angular velocity $\omega$ in  Fig.~\ref{Fig:FreeMotion} shows that the bottle follows clear, convergent and hence, repeatable trends even under the intrinsic variability of the challenger's execution and its final outcome.%}

The dramatic differences between the bottle's kinematics of the control cases (empty, $\phi$ = 0 and full, $\phi$ = 1) and the rest of them build an inescapable proof of the major role of fluid content freedom to move in the challenge.  The best way to analyze what happens with it is by placing ourselves in a frame of reference where the bottle remains fixed as it moves through space. To do so, we follow the centroid of the bottle's rigid-part and rotate it according to its orientation. We show the resulting sequence in Fig.~\ref{Fig:RotReferential}, including, as bottom panels, the corresponding images in the laboratory reference frame. We can easily observe changes in the distribution of water. 
After release (at time $t \approx 330$ ms), a fraction of the water volume rapidly advances to the bottle top while the rest remains at the bottom. After the advancing fraction crosses over the midline, it detaches from the wall ($t \approx 420$ ms) and moves freely through the air until it sloshes against the opposite side of the wall ($t \approx 490$ ms). 
After jet sloshing, water redistributes around, but most of it stays in the same top third of the bottle without significant dynamics. At this stage, water distribution has ended up with two main masses of water at both extremes of the bottle. This has a simple physical explanation: as the bottle is rotating around its center of mass, centrifugal acceleration propels liquid parcels to the farthest volumes available. %An analogous situation occurs with free rotation along the main axis of a partially filled bottle \cite{Andrade_2023}. %; i. e. when rotating a bottle along its central axis: during free fall, water accumulates homogeneously into the walls, respecting the cylindrical symmetry of the bottle. In our case, the bottle has no symmetry on the rotation axis, and the optimal distribution separates water in two volumes.
Finally, when the bottle hits the ground, the upper parcel of water is unable to find any support surface to stop its downward motion. Meanwhile, the bottom parcel is impulsively pushed upward due to the collision with the ground. The two parcels of liquid magically rejoin somewhere halfway inside the bottle and settle in the bottom only later.
%This stage is also visible on a larger time scale in the bottom row of Fig.~\ref{Fig:FlipSequence}. 

\begin{figure}[t!]%[htbp]
\begin{centering}
\includegraphics[width=1\columnwidth]{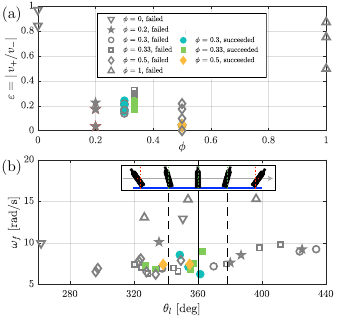}
\par\end{centering}
    \caption{Landing. (a) Restitution coefficient vs. volume filling fraction $\phi$. (b) Angular velocity $\omega_f$ vs. orientation $\theta_l$ just before landing. Gray empty symbols represent failed attempts, while filled and colored ones correspond to successful landings. Symbols vary with the volume filling fraction $\phi$. In panel (b), we highlight angles where the bottle's centroid aligns vertically with the touching point (see green cases in the inset): vertical bottle (\ang{360}, solid line), or inclined at \ang{360}$\pm$\ang{18.4} (dashed lines). Notice that the bottle's motion to the right plays a stabilizing role.\label{Fig:Landing}}
\end{figure}

The rich course of changes in the water distribution shown in Fig.~\ref{Fig:RotReferential} also implies a change in the moment of inertia $I$, which itself
affects $\omega$ through conservation of the angular momentum $L = I\omega$. 
During the free flight, the angular momentum with respect to the center of mass $L$ is conserved. Therefore, the evolution of $L$ is constrained by its initial constant values: $ L  = I (t) \omega (t) =  I_0\omega_0 = L_0$ \cite{DekkerEtAl_2018}.  This law relates the temporal evolution of $\omega (t)$ to that of the mass distribution, quantified by the moment of inertia $I(t)$. Although we do not measure $\omega$ from the center of mass, which includes water, we can still obtain qualitative information. During the early times of free flight, the water spreads around the bottle, simultaneously changing both the center of mass and the moment of inertia. The spreading of water increases $I(t)$ and decreases $\omega(t)$. Later, the jet sloshing further reduces $\omega(t)$ as the jet moves in opposition to the bottle's rotation. %After impact, a more stable water distribution implies stabilization of $\omega(t)$. 
These observations indicate that the water sloshing motion inside the bottle is responsible for slowing down.

%\paragraph{Landing -}
 
The large sample of experiments analyzed, regardless of their outcome, was especially helpful in studying the landing dynamics. We analyzed it by measuring the water's efficiency in attenuating bouncing [Fig.~\ref{Fig:Landing}(a)], and by looking for conditions to succeed (or fail) an upright landing [Fig.~\ref{Fig:Landing}(b)]. 

We characterize the efficiency of water in reducing bounce by introducing the restitution coefficient $\varepsilon = |\, v_+ /v_- |$, where $v_-$ is the vertical velocity of the container before impact, and $v_+$, the one after \cite{Andrade_2023, KillianEtAl_2012, FalconEtAl_1998, GarciaCidEtAl_2015, PachecoVazquez_2013}. 
We computed $\varepsilon$ by tracking the lowest point of the bottle \cite{Andrade_2023}, and we present it in Fig.~\ref{Fig:Landing}(a) as a function of the filling volume fraction. 
We get values of $\varepsilon$ lower than 0.4, meaning the attenuation process at play is very effective. 
The restitution coefficients in the control cases are off the charts  (full bottle, $\phi = 1$ and empty bottle, $\phi = 0$). 
Therefore, fluid motion is responsible for bounce attenuation, as in previous experiments with vertically dropped containers partially filled with liquids \cite{KillianEtAl_2012, Andrade_2023}. In fact, in \cite{Andrade_2023}, we demonstrated that the fluid distributed along the container descends rapidly when the container hits the target, creating a stomp force that keeps the container stuck on the ground. 
Compared to those experiments (\cite{KillianEtAl_2012, Andrade_2023}), we obtain smaller coefficients $\varepsilon$, meaning that the one-turn movement of the bottle is the best in distributing water to produce the stomp force with rapid descent of water: better than natural redistribution after a first straight bounce \cite{KillianEtAl_2012} and better than rapid rotation of the container along its main axis \cite{Andrade_2023}.

The stomp force acts regardless of the final outcome, and thus, what determines the success of the throwing remains intriguing.
To get a hint, Fig.~\ref{Fig:Landing}(b) shows a plot of the orientation $\theta_l$ and the angular velocity $\omega_f$, both at the moment just before landing. Failed landings are presented in gray, while successful ones are shown with filled-colored symbols. Bottles heading nearly vertically to the ground ($\sim \ang{360}$) and with low rotational velocity are more likely to land successfully, as intuitively expected.
Around the vertical, there is a range of stable angles \cite{PahwaEtal_2017, GuEtAl_2021, Cahoon2023}, given by the alignment between bottle's centroid and the touching point ($\Delta \theta = \pm \tan^{-1} (D/h)$), as indicated in the inset of Fig.~\ref{Fig:Landing}(b). The dataset also shows that successful landings asymmetrically occur at angles lower than the criterion, helped by the bottle's rotation inertia.
In general, landings at low rotational velocity are more prone to end upright. This reflects why reducing the angular velocity during the bottle's free-flight motion is key for a successful landing. It also gives the performer control and feedback to improve its technique for future releases.

%\paragraph{Discussion and conclusions -}

To summarize, this letter scrutinizes the bottle flip challenge, focusing on the underlying hydrodynamics, which we identify as key to the trick's feasibility. Its importance is three-fold: First, we show that fast water redistribution after release is responsible for decelerating the bottle flipping during its free-flight motion. Second, we demonstrate that the bottle abruptly attains a nearly steady angular velocity after a jet-like structure of water sloshes the opposite side of the bottle interior. After this, a constant-speed rotation regime in which the fluid content barely moves, i.e., solid-body-like flipping, takes over the dynamics for the rest of the flight.  
Third, the solid-body regime is suddenly over at touchdown when a fluid-mechanical interplay due to massive water motion inside the bottle attenuates any destabilizing bounces at landing.

It is yet surprising that even though fluid dynamics is fundamental along the course of the challenge, the final ingredient for a successful upright landing sounds very familiar to any of us. Challenge achievements rely on simple geometrical stability considerations \cite{PahwaEtal_2017,GuEtAl_2021,Cahoon2023}. After all the flipping, sudden stall, and the unexpected bounceless touchdown, the bottle still needs to land nearly vertically and at sufficiently low angular velocity to stick its landing. 

%In the broader context of fluid dynamics, our system shows that (i) fluid can significantly influence the dynamics of its container, in particular in the presence of jet sloshing. 
%(ii) Fluid could store momentum, both for capturing it to the detriment of its container's motion and to release it appropriately at impact (landing) to have a bounce attenuation effect. 

These observations bring us back to our original question: Why is this mechanical challenge so appealing to the general audience? 
The bottle-flip challenge concatenates mundane, solid-body simple mechanics, which ordinary people master well, with unexpected behavior that seems out of our conventional experience and which fluid is responsible for. The fluid storage of angular and linear momentum allows both the rotation to speed down during free flight and the bounce to attenuate at landing. In practical terms, the complex fluid-solid interaction can be put under the rug, reducing the challenge to the more familiar action of hitting a target with a solid object at the right angle, at reduced speed, and with remarkably lower chances of tumbling. 

%This adds the challenge of a remarkably increased repeatability, which provides repeatability and control that makes it trainable. 

The reduction of angular rotation always occurs, so it does the stomp force at landing. Therefore, there is predictability in the challenge, and one can train to succeed. On the other hand, there is randomness, as the angle before landing should be fine-tuned. This mixture gives the challenge the playfulness of trying something trainable yet hard to achieve. 

%This study was inspired by a fascinating physical phenomenon. Therefore, we close it by comparing it with the equally fascinating falling cat that lands on their feet \cite{Gbur_Cats,CatVideoBBC}. Cats have this impressive ability, regardless of the height or orientation from which they fall. To do so, cats rotate their body differentially, using the conservation of angular momentum to end their fall in a vertical orientation. At landing, they deflect their legs to absorb any bounce, getting a null restitution coefficient \cite{CatVideoBBC}. So, although there are clear similarities, there is one fundamental difference: cats have direct feedback on the situation, and bottles do not. Which ability to land upright is the most remarkable? 

\begin{acknowledgments} 
PG acknowledges Simón Gutiérrez for introducing him to the challenge and helping him with early experiments. We acknowledge Vicente Salinas, Gustavo Castillo, Sergio Rica, David Espíndola, and Adolfo Martínez for thoughtful discussions. This work was funded by the Agencia Nacional de Investigación y Desarrollo (ANID, Chile) through Fondecyt Grants 1221103 (LG) and 11191106 (PG).
\end{acknowledgments}

\end{document}